\documentclass[%
 reprint,
 amsmath,amssymb,
 aps,
]{revtex4-2}

\usepackage{graphicx}% Include figure files
\usepackage{dcolumn}% Align table columns on decimal point
\usepackage{bm}% bold math
\usepackage{braket}
\begin{document}

\preprint{APS/123-QED}

\title{Federated Hierarchical Tensor Networks: a Collaborative Learning Quantum AI-Driven Framework for Healthcare}% Force line breaks with \\

\author{Amandeep Singh Bhatia$^{1}$}
 \email{amandeepbhatia.singh@gmail.com, drasinghbhatia@gmail.com}

\author{David E. Bernal Neira$^{1,2,3}$}%
 \email{dbernaln@purdue.edu}
\affiliation{
$^{1}$Davidson School of Chemical Engineering, Purdue University, West Lafayette, IN, USA
}
\affiliation{$^{2}$Research Institute for Advanced Computer Science, Universities Space Research Association, Mountain View, CA, USA}
\affiliation{$^{3}$Quantum Artificial Intelligence Laboratory, NASA Ames Research Center, Mountain View, CA, USA}

\begin{abstract}
Healthcare industries frequently handle sensitive and proprietary data, and due to strict privacy regulations, they are often reluctant to share data directly.
In today's context, Federated Learning (FL) stands out as a crucial remedy, facilitating the rapid advancement of distributed machine learning while effectively managing critical concerns regarding data privacy and governance.
 The fusion of federated learning and quantum computing represents a groundbreaking interdisciplinary approach with immense potential to revolutionize various industries, from healthcare to finance.
 In this work, we proposed a federated learning framework based on quantum tensor networks, which leverages the principles of many-body quantum physics. Currently, there are no known classical tensor networks implemented in federated settings.
 Furthermore, we investigated the effectiveness and feasibility of the proposed framework by conducting a differential privacy analysis to ensure the security of sensitive data across healthcare institutions.
 Experiments on popular medical image datasets show that the federated quantum tensor network model achieved a mean receiver-operator characteristic area under the curve (ROC-AUC) between 0.91-0.98. 
 Experimental results demonstrate that the quantum federated global model, consisting of highly entangled tensor network structures, showed better generalization and robustness and achieved higher testing accuracy, surpassing the performance of locally trained clients under unbalanced data distributions among healthcare institutions.

\end{abstract}

\maketitle

%\tableofcontents
\section{INTRODUCTION}
Advancing medical research through collaborative healthcare data sharing presents challenges stemming from data heterogeneity (in terms of formats, standards, and structures) and stringent privacy regulations. 
It is widely acknowledged that an ample dataset is essential for machine learning algorithms to grasp robust patterns and exhibit effective generalization across various medical domains, from drug discovery to disease diagnosis.
 The complexity and scale (i.e., millions or billions of parameters) of deep learning models and the need for substantial data enable them to automatically learn intricate representations and relationships in the data, leading to their remarkable performance in various domains \cite{1}.
 Healthcare institutions are bound by stringent privacy standards as patient data is laden with sensitive and personally identifiable information.
 There is apprehension that data sharing could jeopardize patient privacy, potentially resulting in legal and ethical consequences.

\begin{figure*}[!ht]
	\centering
	\includegraphics[scale=0.8]{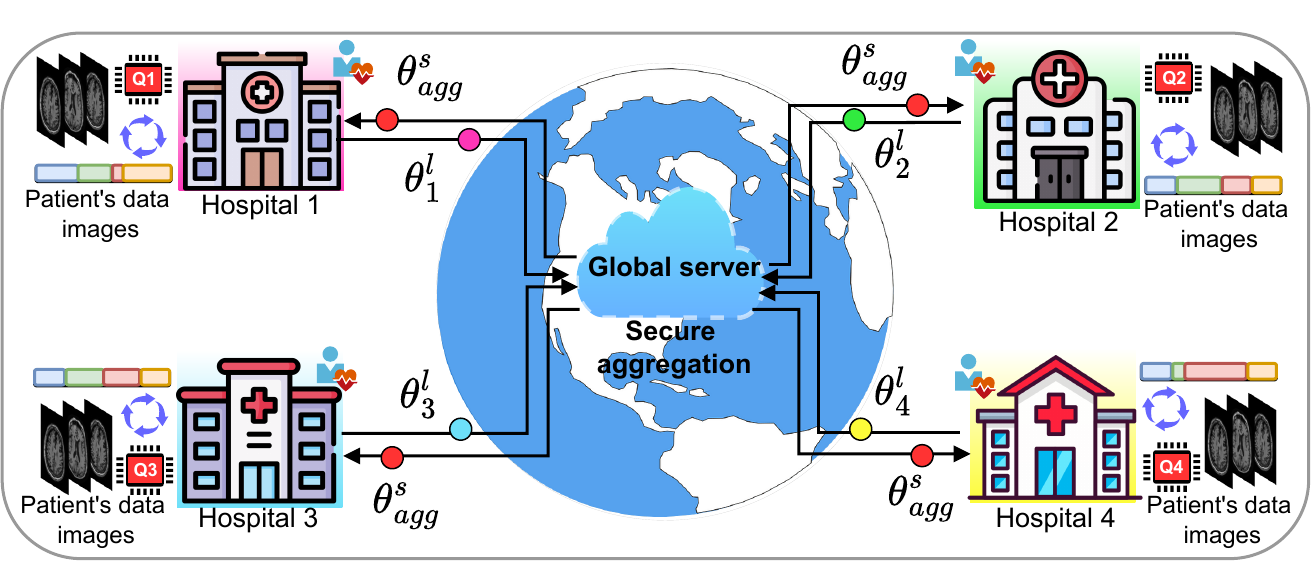}
	\caption{\textbf{Collaborative Quantum AI-based framework for healthcare advancement:} Multiple hospitals participate in Quantum Federated Learning by collaboratively training models while preserving patient's privacy. Step 1: Hospitals locally process patient's image data using the quantum tensor networks (QTNs) model. Step 2: The quantum model updates of each hospital, not the raw data, are shared with the global server. Step 3: The global server aggregates the model updates from all participating hospitals and sends the updated state of the model back to all hospitals, ensuring that each participant benefits from the collective knowledge gained during the training process.    }
 \label{framework}
\end{figure*}

The concept of federated learning (FL) holds great promise in advancing medical research by enabling collaborative analysis of decentralized data and addressing challenges related to data heterogeneity, communication, security, technical complexity, and legal considerations.
 Initially, FL was designed for various domains, including edge and mobile device use cases \cite{2}, but it has recently gained momentum in the realm of healthcare applications.
 It allows model training on edge nodes, enabling real-time insights and decision-making without the need for extensive data transfers to a central server.
 Only the model updates are sent back to a central server, where they are aggregated to improve the global model, as shown in Fig.~\ref{framework}.
 It enables the global model to gain insight from various data sources while maintaining privacy and reducing data transfer costs, especially when the decentralized training is scaled with multiple nodes.
 In Fig.~\ref{framework}, the global server is at the heart of federated learning. 
 It receives updates from all the hospitals and aggregates them to create a new global model. Overall, it acts as a central aggregator and coordinates the learning process to collectively enhance the knowledge base of disparate entities without sharing data.

Recent strides in federated learning empower the construction of complex classical machine learning models trained across decentralized systems. The effective integration of FL in healthcare stands poised to significantly transform precision medicine on a broad scale and can improve clinical workflow, while also upholding stringent standards of privacy and governance \cite{3}. Noteworthy examples illuminate the promising results of federated learning algorithms in healthcare applications including detecting rare cancer disease of glioblastoma across six continents \cite{4}, identifying brain anomalies across six institutions \cite{5}, integrating differential privacy to clinical and epidemiological research \cite{6}, safeguarding privacy in medical imaging \cite{7}, conducting a multi-national study to detect COVID-19 lung abnormalities in CT scans \cite{8}, utilizing FL deep generative models for healthcare \cite{9}, ensuring privacy in medical image analysis \cite{10}, diagnosing COVID-19 with collaborative FL \cite{11}, and numerous other breakthroughs. FL stands out as the foremost privacy preservation technique of the next generation, gaining substantial traction across industries and notably within medical AI applications.

In recent years, quantum computing, especially distributed quantum computing, including quantum machine learning (QML), has made remarkable progress \cite{24}.
 Its ability to leverage the combined power of distributed quantum resources surpasses the constraints of individual quantum nodes \cite{48}. 
 QML has shown the ability to profoundly influence diverse domains in various practical applications, ranging from scientific research and engineering to business and finance. 
 Due to parallelization and scalability, distributed quantum computing has shown a significant potential to leverage the combined strength of quantum nodes, potentially overcoming the limitations associated with individual quantum nodes. Coupled with theoretical assurances, some preliminary findings have demonstrated its computational edge compared to traditional machine learning tasks such as classification \cite{12, 13}.
 Huang et al. \cite{14} assessed the potential quantum advantage in learning tasks and subsequently conducted a comparative analysis of classical and quantum machine learning models regarding their ability to predict outcomes in physical experiments \cite{15}.

Several references in the relevant literature in this area have explored novel quantum algorithms to optimize global models using decentralized data sources.
Some notable examples of quantum machine learning algorithms in federated settings include quantum neural networks with pre-trained classical models (hybrid) \cite{16}, quantum neural networks \cite{17}, variational quantum circuits, variational quantum tensor networks \cite{18}, quanvolutional neural networks \cite{19},  quantum convolutional neural networks \cite{20}, and handling privacy sensitive clinical data with federated quantum machine learning \cite{50}. 
Some previous studies have demonstrated the resilience of quantum federated model parameters against eavesdropping \cite{21}, performed quantum federated learning with blind quantum computing \cite{22}, differential privacy \cite{23}, and have addressed privacy-preserving concerns with gradient hiding \cite{24}. 

\begin{figure*}[!ht]
	\centering
	\includegraphics[scale=0.58]{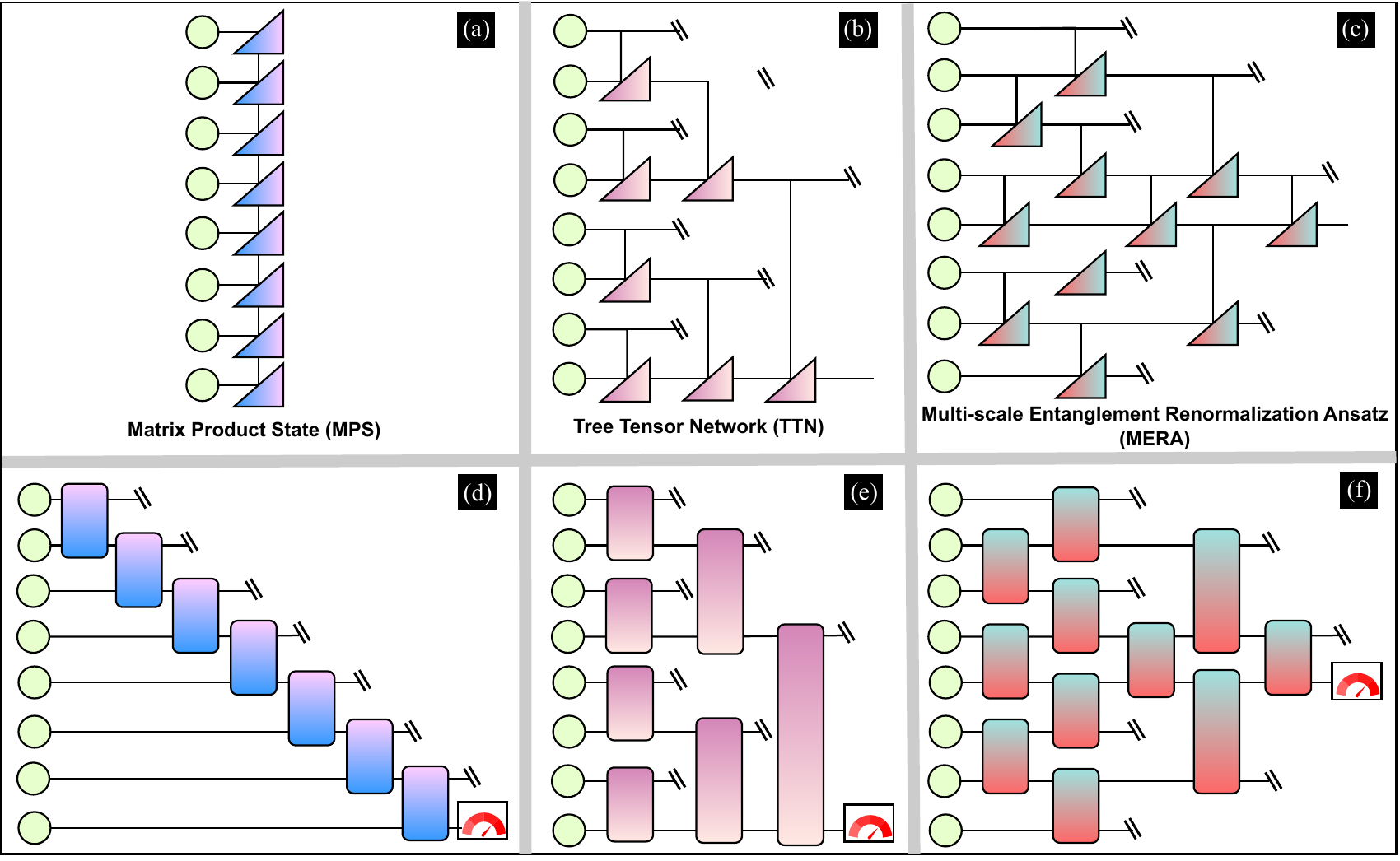}
	\caption{\textbf{ Classical tensor networks and their quantum circuit representation counterparts.} (a) Matrix product state (MPS): A one-dimensional chain of tensors that captures one-dimensional quantum systems efficiently. (b) Tree tensor network (TTN): It extends the idea of MPS to higher dimensions, using a tree-like structure to capture entanglement across multiple dimensions. (c) Multiscale entanglement renormalization ansatz (MERA): It captures logarithmic correlations while scaling with the length of 1D. (d-f) Implementation of the model as a quantum circuit. The tensor elements within the tensor networks are substituted with unitary operations to construct quantum circuits. Circles indicate eight qubit inputs prepared in a product state (left-hand side of the circuit), hash marks denote qubits that have not been observed beyond a specific stage in the circuit, and qubits are entangled via two-qubit unitaries (in square blocks). A preselected qubit is sampled (in red measurement operator), and the resulting distribution is considered as an output of the quantum model.}
 \label{tensornetworks}
\end{figure*}

In parallel, tensor networks (TNs) are considered promising candidates for quantum machine learning (QML) architectures \cite{46}. 
TNs provide a compact and efficient representation of quantum states, using tensor structures that encapsulate substantial amounts of correlated information. 
TNs enable the application of individual local operations on each tensor node, eliminating the need to compute the entire tensor. 
Tensors can be combined through contraction over connected indices or broken down into multiple interconnected tensors. 
For a more in-depth technical exploration of TNs, readers can delve into a comprehensive introduction \cite{25} and review articles that focus on specific architectures \cite{26, 27, 28}.
 Matrix Product State (MPS) is the simplest mathematical framework for describing quantum states in many-body systems \cite{30, 33}. 
The key idea behind MPS is that the entanglement in a many-body quantum state is often localized, meaning that each tensor is primarily entangled with its neighbors.
 It is a widely studied one-dimensional TN layout in which the level of entanglement remains constant. 
In hierarchical tensor networks (HTNs), input or output tensors are pooled across several internal layers rather than directly interconnected.
 The tree tensor network (TTN), a specific form of HTN, is designed to efficiently represent high-dimensional tensors by decomposing them into a sequence of lower-dimensional tensors interconnected hierarchically.
 One of the key advantages of TTN is its ability to capture long-range correlations in high-dimensional data while maintaining a low computational cost \cite{46}.
 Another more complex TN layout is the multiscale entanglement renormalization ansatz (MERA) that goes beyond TTN in capturing symmetry and handling higher entanglement at different length scales, making it a valuable tool for studying complex quantum systems \cite{29, 46}, as shown in Fig.~\ref{tensornetworks}(a-c).

The quantum-inspired formulation of TNs is straightforward, and QTNs can be achieved through multi-qubit gates, where incoming and outgoing qubits represent the bonds of the tensor node. 
A crucial advancement in integrating tensor networks (TNs) into quantum computers was the formulation of a breakdown technique enabling the transformation of multi-qubit nodes into two-qubit unitaries with exceptional fidelity.
 This breakthrough facilitates the efficient implementation on NISQ devices.  
 Due to the direct similarity, models and data from classical tensor networks can be seamlessly transformed into their quantum variants, and vice versa \cite{46}.

Taking into account the current development of quantum computer hardware, QTNs find applications in various machine learning tasks, including classification \cite{29, 30, 31, 32},  generative TNs \cite{33}, entanglement-based feature extraction \cite{34}, and compression \cite{35}. 
Parameterized circuits with a simple gate set are used for MPS, TTN, and MERA, which can be implemented on NISQ devices efficiently, as shown in Fig.~\ref{tensornetworks}(d-f). 
The quantum circuit starts by applying a set of two-qubit unitaries to the input and leaving one of the qubits unobserved (i.e., indicated by hash marks).
 The optimization of parameters in quantum tensor networks can be achieved using the stochastic gradient descent (SGD) method. 
Quantum machine learning models based on tensor networks act as linear classifiers operating within a feature space that expands exponentially in the number of data qubits \cite{49}. 

\begin{figure*}[!ht]
	\centering
	\includegraphics[scale=0.75]{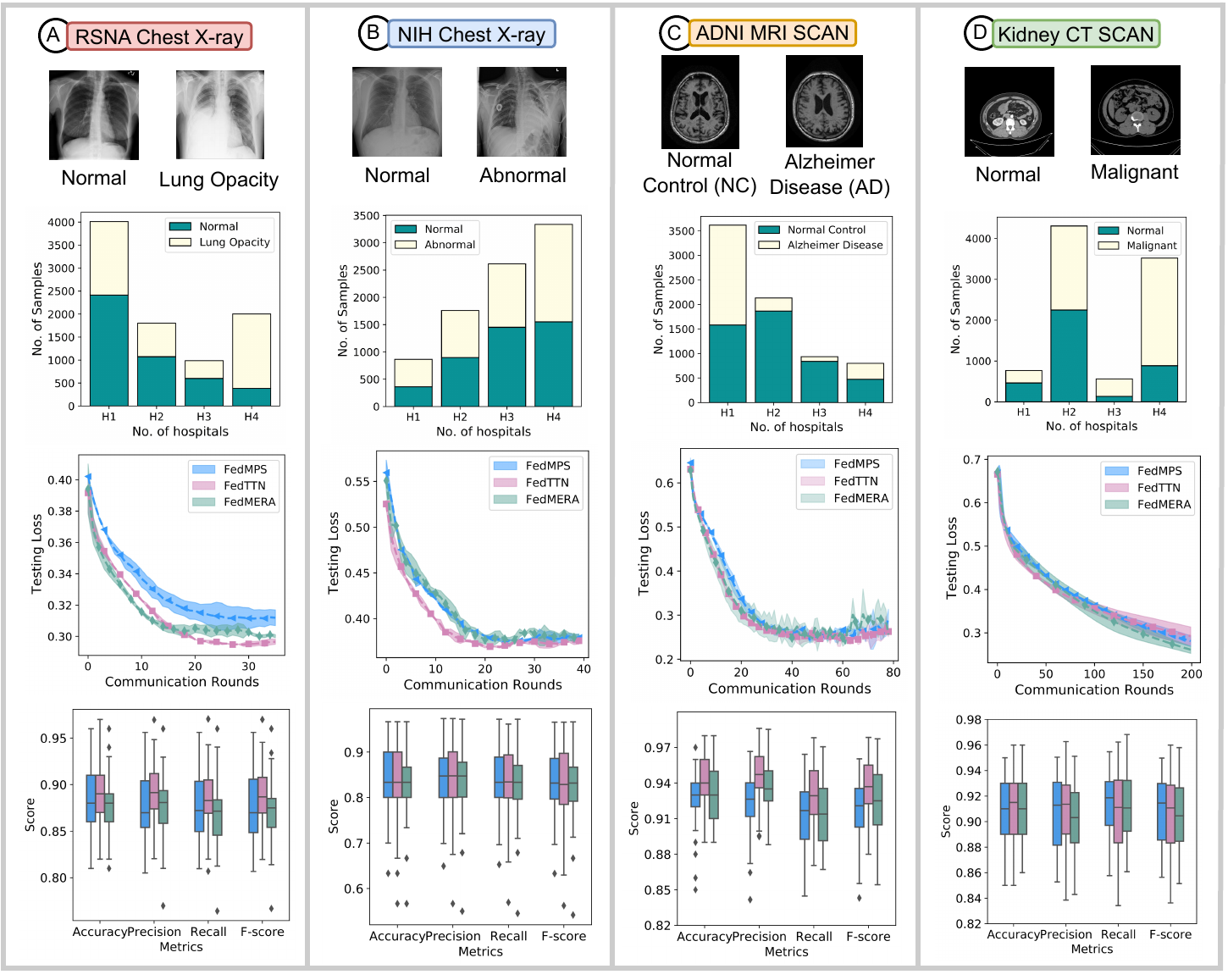}
	\caption{\textbf{Performance of federated quantum tensor networks on different datasets.} (a) RSNA chest X-ray dataset: (b) NIH chest X-ray dataset: (c) ADNI MRI-scan dataset: (d) Kidney CT-scan dataset. An example of medical images of each dataset is displayed in the first row. The unequal distribution of all datasets among four hospitals/clients/participants is provided in the second row. The testing loss curves of three federated QTN models (FedMPS, FedTTN, and FedMERA) are presented in the third row. The illustration of the box plots shows the performance metrics: testing accuracy, precision, recall, and f-score of federated QTNs across all datasets.}
  \label{mainresults}
\end{figure*}

\subsection{Current challenges}
 Today, several healthcare institutions hesitate to share data and participate in big data initiatives due to their data's sensitive nature. 
Healthcare data is subject to strict regulations and its sharing for research purposes could potentially breach privacy laws. 
As a result, data sharing and privacy concerns hinder collaboration for research and model development. 
One of the common challenges in healthcare is that medical data is dispersed among various institutions and systems, making it difficult for centralized machine learning to use diverse data sources effectively. 
In some cases, data unavailability, incompleteness, or bias poses obstacles to advancing machine learning solutions in the healthcare sector.  
One major challenge in tensor network theory is the computational efficiency and interpretability trade-offs. 
Tensor networks, by their nature, can capture complex relationships within data. However, this strength comes at the cost of reduced interpretability. 

\subsection{Motivation}
 Considering that tensor networks can serve as representations for both neural networks and quantum circuits, it becomes a natural objective to investigate the intersection of these two fields in federated settings. The main motivation is to unlock the potential benefits of quantum tensor networks and federated learning in the healthcare sector. This includes creating improved diagnostic tools, improving medical image analysis, and ensuring meaningful results for rare diseases when smaller healthcare institutions lack sufficient data to train an accurate predictive model. 

An advantage of using quantum tensor networks is that tensor contraction, which can be computationally expensive on classical hardware, can be naturally executed as part of the quantum circuit \cite{46}. 
Motivated by the efficient training of quantum tensor circuits and the expressive power of hierarchical structures with exponentially fewer parameters, a quantum AI federated framework can be created to collaboratively assist the healthcare sector. 

The principles of entanglement entropy's area laws offer inherent insights into the interpretation or computational capabilities of tensor networks in simulating quantum systems \cite{47}. 
Integrating tensor networks in quantum machine learning not only significantly improves interpretability but also effectively tackles challenges related to high-dimensional spaces \cite{46}. 
Various tensor network architectures offer the flexibility to represent large tensors or wave functions. 
These architectures provide benefits in terms of universality and interpretability.
Implementing such an innovative framework in the healthcare industry can contribute to improving healthcare outcomes.

\subsection{Main contributions}
  To summarize, this paper makes the following contributions:

\begin{itemize}
    \item  Proposal of a federated learning framework based on quantum tensor networks, or FedQTN, for collaborative learning between multiple healthcare institutions, enabling QTNs to train on nonindependent and identically distributed medical images.
    
    \item  Demonstration of the learning capacity of FedQTNs on popular medical image classification datasets, validating our approach and its adaptability to heterogeneous data. 
    This includes chest radiographs from the RSNA and NIH datasets, MRI brain scans from the Alzheimer's ADNI dataset, and CT scans from the CT-kidney dataset.
    
    \item Evaluation of the strength and versatility of various QTNs that involve the integration of differential privacy mechanisms that preserve privacy in federated settings.
    This mechanism safeguards the security of medical data across healthcare institutions.

    \item Observations that quantum tensor networks with higher entanglement among qubits, such as TTN or MERA, exhibit higher classification test accuracy than MPS. Furthermore, TTN stands out for its faster training convergence speed compared to all. 
    
\end{itemize}

The remainder of the article is organized as follows. Section 2 includes the experimental results on various medical datasets. Section 3 provides a discussion of the findings and outcomes. Finally, Section 4 presents the dataset settings and a description of the federated quantum tensor networks.

\begin{figure*}[!ht]
	\centering
	\includegraphics[scale=0.62]{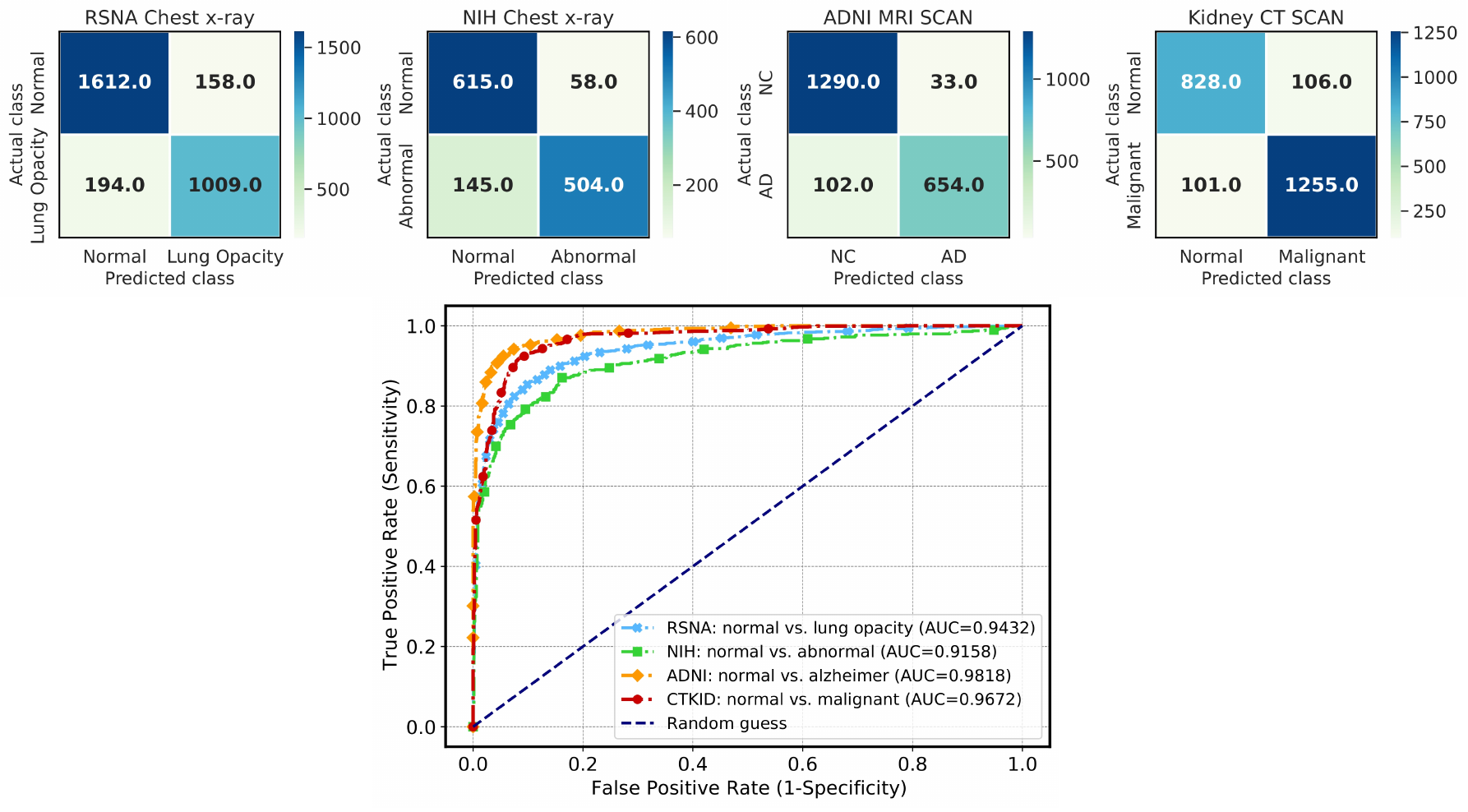}
	\caption{\textbf{Confusion matrices and receiver operating characteristic curves (ROCs) of federated TTN model performance on different datasets}. Left: RSNA chest X-ray for normal versus lung opacity classification. Middle: NIH chest X-ray for normal versus abnormal classification. ADNI MRI-scan for normal cognitive versus Alzheimer's disease (AD) classification. Right: Kidney CT-scan for normal versus malignant classification. ROCs depict the performance of quantum TTN on various datasets in federated settings. The maximum Area Under the Curve (AUC) reached 0.9818 for ADNI MRI-scans, while the minimum AUC was 0.9158 for the NIH chest X-ray dataset. }  
 \label{aucplot}
\end{figure*}

\section{RESULTS}
\noindent \textbf{Model performance on the RSNA chest X-ray dataset}

We first tested the recognition capacity of chest radiographs (normal = 8851, 59.4\% versus lung opacity = 6012, 41.6\%) of federated quantum tensor networks (FedQTNs) in the chest radiograph data set RSNA \cite{37} for performance evaluation.
 In total, 14,863 samples are adopted for classification evaluation, where 70\%, 10\%, and 20\% of the samples are used for training, validation, and testing, respectively.
 An original image size of 512$\times$512 is too large for current quantum hardware. 
 To efficiently simulate all circuits, each input image size is downscaled to 32$\times$32,  leading to the loss of significant information that could potentially be valuable for classification. In general, classical machine learning demonstrates strong performance under conditions of ample training data. 
 We investigated the robustness and stability of FedQTNs when distributing the data unevenly among the four hospitals/participants/clients (H1, H2, H3, H4), as shown in Fig.~\ref{mainresults}(a). 
 Hospital H1 has the highest number of training samples, while H3 has the fewest. 
 The test dataset contains 2973 radiographs (normal=1770, 59.6\%; lung opacity=1203, 41.4\%). 
 As the testing loss is widely used to measure the quality of training capability, we consider this metric versus training communication rounds for FedMPS, FedTTN, and FedMERA models. 
 The FedTTN model outperforms other federated models with its quicker training convergence and higher test accuracy, as shown in Fig~\ref{mainresults} (a).
 One-dimensional FedMPS achieves higher testing loss than hierarchical FedQTN models. FedTTN achieves an AUC of 0.9432, classifying normal and lung opacity chest radiographs. 
 The confusion matrix and ROC of the FedTTN model are shown in Fig.~\ref{aucplot}.

\begin{table*}
\caption{Summary of datasets}
\begin{tabular}{ |p{3.5cm}|p{1.4cm} p{2.0cm} |p{1.7cm} p{1.7cm}| p{1.7cm} p{1.7cm}|}
\hline

\multicolumn{1}{|c|}{} & \multicolumn{2}{|c|}{RSNA chest X-ray} & \multicolumn{2}{|c|}{NIH chest X-ray} & \multicolumn{2}{|c|}{Kidney CT-scan} \\
\hline
Train-Test-Split & Normal & Lung opacity & Normal & Abnormal & Normal & Malignant \\
\hline
Train subjects (Image\#)	&	7989 	&	4328 &	4261	&	4313	&	3736 &	5422\\
Test Subjects (Image\#) &	1770	&	1203 &	663	&	649	&	934	&	1356	\\
\hline
\end{tabular}
\label{table_summary}
\end{table*}

\noindent \textbf{Model performance on the NIH chest X-ray dataset}

In this section, we applied QTN models on the NIH chest X-ray dataset \cite{38} to differentiate between frontal-view normal and abnormal chest radiographs in federated settings. 
A total of 11,574 radiographs were acquired. Of these, 10,252 were divided into training (9896) and validation (356) datasets, while the remaining 1,322 images (normal=663 versus abnormal=649) were reserved for testing. 
An original image size of 1024$\times$1024 was reduced to 32$\times$32, which can have a detrimental impact on the model's accuracy. 
The training dataset distribution of four hospitals is shown in Fig.~\ref{mainresults}(b). 
Box plots show that FedQTNs achieve similar testing accuracy, precision, recall, and f-score. 
However, the FedTTN converged better than other QTN models. 
Fig.~\ref{mainresults}(b) illustrates the testing loss curves of three FedQTN models.
 Despite the challenges posed by inadequate and unbalanced hospital training data, together with a substantial reduction in image size, the FedQTN model achieved an AUC of 0.9158, with a specificity of 91.38\% and a sensitivity of 77.65\%. 
Fig.~\ref{aucplot} presents the ROCs and confusion matrix of the FedTTN model.

\noindent \textbf{Model performance on the ADNI MRI-scan dataset}

Next, we evaluated the performance of FedQTNs in differentiating individuals with normal control (NC) from those diagnosed with Alzheimer's disease (AD) based on magnetic resonance imaging (MRI) scans. 
For the Alzheimer’s Disease Neuroimaging Initiative (ADNI) MRI-scan dataset \cite{39}, a total of 331 subjects (normal cognition=179 and Alzheimer's disease=152) were collected. 
Axial 2D slices were generated from 3D T1-weighted MRI brain images and then downscaled to 32$\times$32 from the original size of 160$\times$192. 
A total of 10,231 MRI scans were collected. Out of these,  the training set consists of 7,482 and validation 670 samples, while the remaining 2,079 images (NC=1,323 versus AD=656) were used for testing.
 The sample of images is shown in Fig.~\ref{mainresults}(c). 
Furthermore, the data set is distributed randomly among four hospitals, where H1 has the maximum number of samples, i.e., 48\% of the training dataset, and H3 has only 12\% of training samples. 
We have an imbalanced class distribution among hospitals.

The performance of individual FedQTN global models is reflected through testing loss against the training communication rounds.
We observed that the hierarchical FedTTN and FedMERA exhibit a comparable convergence initially; however, FedMPS and FedMERA encounter fluctuations in loss at later stages, as shown in Fig.~\ref{mainresults}(c). 
Similar to previous results, FedTTN outperformed the other federated quantum models and achieved higher testing accuracy, precision, recall, and f-score after training, as illustrated in Fig.~\ref{mainresults}(c). 
The confusion matrix for the ADNI MRI-scan test set in FedTTN is shown in  Fig.~\ref{aucplot}. 
An AUC of 0.9818  was achieved, with a specificity of 97.50\% and a sensitivity of 86.5\%.
The positive predictive value attained 95.19\%, and the negative predictive value was 92.67\%. 
These results suggest that FedQTN models are proficient in distinguishing between normal MRI scans and those identified with Alzheimer's disease,  irrespective of the uneven data distribution among the participants.

\begin{figure*}[!ht]
	\centering
	\includegraphics[scale=0.53]{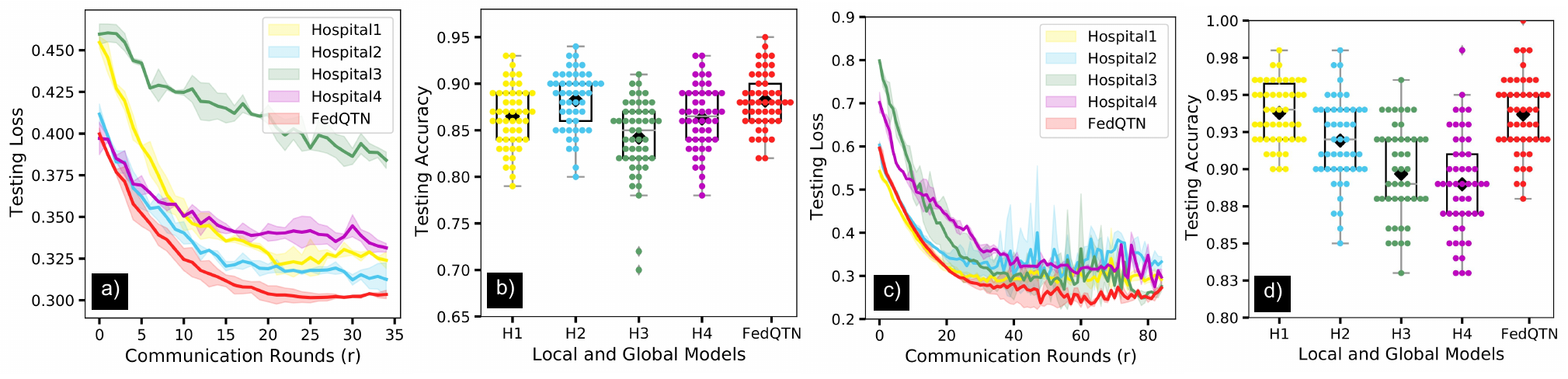}
	\caption{\textbf{Performance of individual local models and the global model.} (a) Testing loss curves for the four locally trained hospitals on RSNA chest radiographs and global FedQTN. (b) Testing accuracy performance across four hospitals and FedQTN is illustrated through box plots, revealing the impact of insufficient data distribution (c-d) Testing loss curves for the four locally trained hospitals on ADNI MRI scans. The global FedQTN model significantly outperforms the locally trained models, achieving a testing accuracy of 94.5\% along with a smoother convergence. }  
 \label{client_results_plot}
\end{figure*}
\noindent \textbf{Model performance on the Kidney CT-scan dataset}

\begin{table}
\caption{Summary of demographics of ADNI dataset \cite{39}}
\begin{tabular}{ |c|c|c|c|c|c| }
\hline
 & \multicolumn{2}{|c|}{Sex} & \multicolumn{2}{|c|}{Age} &  \\
\hline
\centering
 Diagnosis & M &	F &	Mean & Std	&	\# of Subjects\\
 \hline
 Alzheimer's disease (AD) & 77  & 75 & 76.57 &  7.47 & 152\\
 Cognitive normal (CN) &  84 & 95 & 77.40 & 5.38 & 179\\
 \hline
\end{tabular}
\label{table_adni}
\end{table}

In this section, we examined the performance of FedQTNs for the diagnosis of kidney diseases.
A CT-kidney dataset comprising 12,446 CT-scan images of the kidney is collected, featuring classes: Cyst, Tumor, and Stone \cite{40}. 
These abnormalities were binned into the “malignant” category, and negative studies were included in the “normal” category. 
After categorizing into “malignant” and “normal”, each image is reduced to a uniform 512$\times$512 size. 
A total of 10,156 CT-scans were separated into training (normal=3736, 40.8\% versus malignant=5422, 59.2\%) and validation sets and 2290 (normal=934, 40.8\% versus malignant=1356, 59.2\%) AUC testing purposes. 
Further, each image was reduced to an 18$\times$18 (i.e., 324 components) using Principal Component Analysis (PCA) while keeping about 99\$ of variance and preserving as much variability in the original dataset. 
Next, we divided the whole image into 2$\times$2 patches and encoded it into a 4-qubit quantum state. Here, we used a simple unitary block as a two-qubit unitary. 
Fig~\ref{mainresults}(d) illustrates the distribution of normal and malignant images across four hospitals. Notably, hospital H2 accounts for around 47\% of the total samples, while H3 represents a mere 6\% of the training dataset. 
We find that all FedQTNs demonstrate smoother convergence and strong generalization capabilities and attain an impressive testing accuracy of 91.5\%, as shown in Fig.~\ref{mainresults}(d). 
The confusion matrix of binary label recognition is provided in Fig.~\ref{aucplot}. 
FedTTN model achieved an AUC of 0.9672, classifying normal and malignant kidney CT scans with a specificity of 88.6\% and sensitivity of  92.5\%.

\section{DISCUSSION}
In this paper, we have proposed a federated learning framework based on quantum tensor networks (FedQTNs), integrating quantum computing and collaborative learning in the domain of healthcare intelligence. 
FedQTN enables healthcare institutions to collaborate without sharing sensitive patient data and refine a globally reliable model in the presence of statistical heterogeneity.
 Each institution locally trains QTN models on its own dataset and shares model updates during the collaborative learning process. 

Federated learning has recently found applications in diverse fields like edge computing, such as smartphones, tablets, wearables (smartwatches and fitness trackers), Internet of Things (IoT) devices, and other sensor-equipped gadgets. 
However, the context of medical imaging introduces additional complexity and presents distinctive challenges, such as dealing with imbalanced cohort sizes and high-dimensional data.
 To the best of our knowledge, no published studies have implemented or evaluated the utility of quantum tensor networks in federated settings.

Our experimental observations have shown that quantum federated learning improved generalization performance over different medical datasets, reflecting effective optimization with imbalanced training data distributions across healthcare institutions.
 As noted in our results, the use of FedQTNs significantly improved the performance of local quantum models within healthcare institutions, especially when training with uneven cohorts (e.g., hospitals H3 and H4 have mild cases of normal patients in ADNI dataset and H3 comprises only 8\% of training samples from the RSNA dataset). 
 FedTTN model achieved a higher true-positive rate and a significantly reduced false-positive rate, as shown in the confusion matrices and the receiver operating characteristic (ROC) plots (Fig.~\ref{mainresults} and Fig.~\ref{aucplot}). 
 Significantly, in federated settings, the hierarchical quantum tensor network models (TTN and MERA) exhibited a notable improvement in generalizability compared to the one-dimensional MPS model.
 Moreover, the performance of individual locally trained models is compared with the global model for RSNA and ADNI datasets in Fig~\ref{client_results_plot}(a-d), respectively. 
 It has been demonstrated that QTN-based federated learning effectively achieves collaborative optimization with nonindependent and identically distributed medical images among hospitals. 
 We observed that hospitals with a limited number of training medical images exhibited higher testing loss, resulting in reduced testing accuracy. 
 The FedQTN model effectively addresses this issue. 
 As shown in Fig.~\ref{client_results_plot}(d), the accuracies of locally trained models for H3 and H4 has been improved from 89.2\% and 88.5\% to 94.5\%. 
 The global FedQTN model significantly outperforms locally trained models, attaining higher accuracy and demonstrating faster and more seamless convergence. 
 Our strategy based on quantum tensor networks provides several benefits. 
 It shows that a quantum AI-driven federated learning framework could potentially achieve clinical standards for accurate diagnosis, even in scenarios with imbalanced data distributions.

\begin{figure}[!ht]
	\centering
	\includegraphics[scale=0.62]{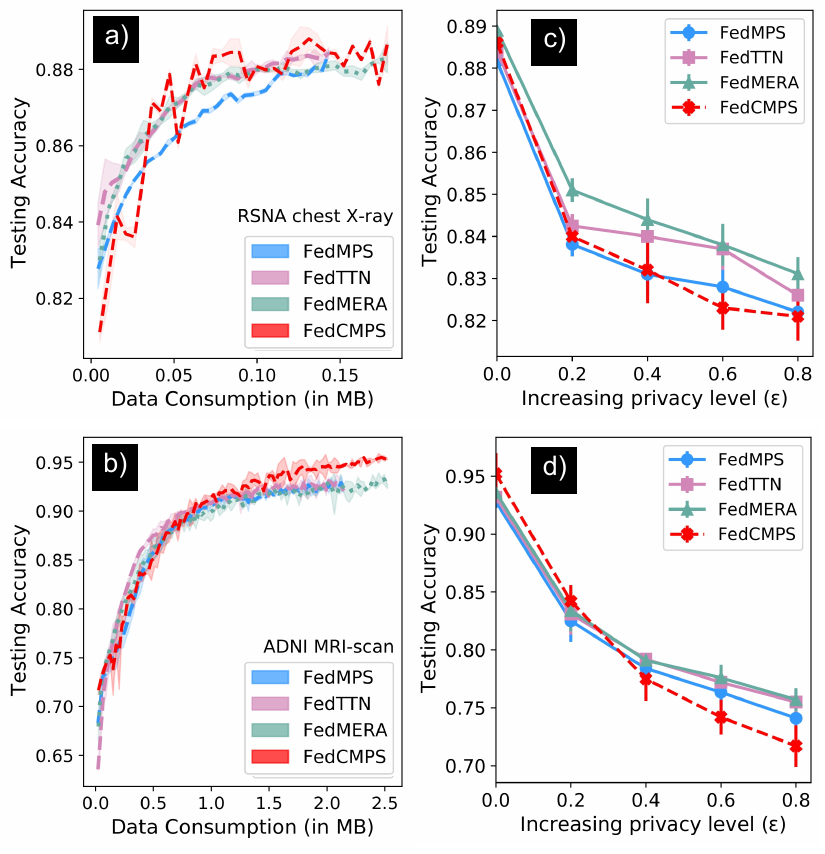}
	\caption{\textbf{Communication overhead and robustness against local differential privacy.} (a-b) Comparing communication efficiency of QTNs and classical MPS in federated settings for RSNA and ADNI datasets, respectively. It shows the total message size (in megabytes) exchanged between four hospitals and the global model. (c-d) Robustness of FedQTNs and FedCMPS in terms of testing accuracy against the local differential privacy noise. The higher the noise ($\epsilon$), the more privacy we can get. The performance of classical MPS decreases as the noise level increases significantly compared to hierarchical FedQTN models. }
 \label{noise_results}
\end{figure}

Next, we analyze the communication overhead of all federated QTN models and compare them with federated classical MPS on RSNA and ADNI datasets. 
In experiments with four hospitals using the RSNA data set, FedTTN and FedMERA demonstrated efficient performance by exchanging only 0.2 MB of data and requiring fewer communication rounds to achieve commendable results on the test data compared to FedMPS. 
On the other hand, the federated classical MPS exchanged 0.2 MB of data between the local models of hospitals and the global server, but exhibited noticeable fluctuations throughout the communication rounds, as shown in Fig.~\ref{noise_results}(a). 
For the ADNI MRI-scan dataset, the testing accuracy of all FedQTNs is equivalent, and 2.5 MB of data is exchanged. 
FedCMPS showed marginally better accuracy on the test set, as depicted in Fig.~\ref{noise_results}(b).

In addition, we examine the stability and robustness of FedQTNs with local differential privacy at the client level. 
This process involves two key steps: initially, we clipped random samples from local training data from each hospital using a clipping threshold (\textit{C}=1.0); subsequently, constant noise ($\epsilon$) was added to the clipped gradients, introducing heterogeneity between hospitals/clients. 
Each hospital uses differentially private ADAM (DPAdam) when training the local QTN model. 
As illustrated in Fig.~\ref{noise_results}(c,d), the testing accuracies of FedQTN outperformed those of FedMPS and FedCMPS with DPAdam gradient clipping and noise addition. With no privacy constraint ($\epsilon$=0.0), the FedCMPS is marginally better than FedQTNs on the ADNI dataset. 
We observed that hierarchical QTNs show a lower decrease in performance than other methods as the level of noise increases.

\section{METHODS}
In this section, we provide an in-depth introduction to our federated learning framework based on quantum tensor networks. 
We first define the problem studied in this paper and then delve into the details of our approach.

\begin{figure}[!ht]
	\centering
	\includegraphics[scale=0.7]{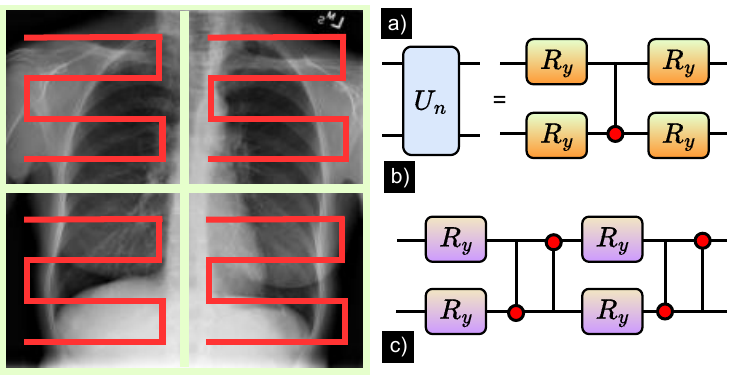}
	\caption{\textbf{Encoding two-dimensional data and two-qubit unitaries. } (a) Patch-based encoding: The whole medical image is split into patches. Each patch is encoded into a quantum state using single-qubit rotations. Subsequently, these quantum states are individually fed into QTNs and the results are concatenated. (b) Simple unitary block: Two single-qubit $R_y$ rotations, followed by CNOT and additional $R_y$ rotations. (c) Strongly entangling block: Each layer constitutes two single-qubit $R_y$ rotations, followed by a forward CNOT and a reverse CNOT operation. We used a strongly entangling block for all datasets except for the Kidney CT-scan dataset, where a simple unitary block was used. }
 \label{encoding2}
\end{figure}

\subsection{Study population}
 We studied four different databases. 
1. Data set from the Radiological Society of North America (RSNA) and Society of Thoracic Radiology (STR) pneumonia detection challenge \cite{37}. 
The RSNA chest radiograph data set contains chest radiographs classified as normal, pneumonia-like lung opacity, and abnormal without lung opacity. 
The term “pneumonia-like lung opacity" encompasses observations such as pneumonia, consolidation, and infiltration. Our evaluations focused on distinguishing between normal (n=8,851) and abnormal (or pneumonia-like lung opacity) (n=6,012) chest radiographs. 
2. National Institutes of Health database: NIH released the ChestX-ray dataset in 2017, which comprises 112,120 frontal radiographs from 30,805 distinct patients \cite{38}. 
Tang et al. \cite{41} used a natural processing (NLP) tool and categorized abnormalities (consolidation, edema,  effusion, emphysema, fibrosis, hernia, infiltrate, mass, nodule, pneumonia, pneumothorax, pulmonary atelectasis, pleural, and thickening) into the “abnormal" class.
 In contrast, studies without abnormalities were placed in the “normal" category \cite{42}. 
A total of 11,574 radiographs were acquired. Of these, 9,896 (normal = 4,261 versus abnormal = 4,313) were used for training, and 1,322 images (normal=663 versus abnormal=649) were used for testing.
3. Alzheimer’s Disease Neuroimaging Initiative (ADNI) MRI-scan dataset \cite {39}:  
We collected demographics, medical history and brain magnetic resonance imaging (MRI) scans. 
A total of 331 subjects (normal cognition = 179 and Alzheimer's disease = 1502), as shown in Table~\ref{table_adni}.
 We extracted the relevant 2D slices (axial, one, or multiple per subject in interesting areas). 
A total of 10,231 MRI-scans were collected, comprising 6,555 participants with normal cognition (NC) and 3,676 participants with Alzheimer’s disease (AD). 
4. Kidney CT-scan dataset: A CT-Kidney dataset is publicly available \cite{40}. It consists of 12,446 annotated CT-scans categorized into four classes: Normal (n=5,077), Cyst (n=3,709), Stone (n=1,377), and Tumor (n=2,283).
 We grouped the abnormalities (cyst, stone, and tumor) into a malignant class and classified the labels as normal versus malignant.
 The summary of datasets is given in Table~\ref{table_summary}. 

\subsection{Implementation details}
All methods use the Adaptive Moment Estimation (Adam) optimizer; the learning rate of global server and local models is set to 0.001, mini-batch size of 8, and weight decay of 1e-4. 
All the experiments are performed using a quantum programming framework, TensorFlow Quantum (TFQ), by Google \cite{43}. 
To implement classical MPS, we used the Tensor Network Library \cite{44}.
For differential privacy, we adopted a differentially private DPAdam optimizer, clipping \textit{C} = 1.0, and noise level $\epsilon=[0.2, 0.4, 0.6, 0.8]$. 
The hardware and software utilized in the paper are described within the section.~\ref{details}

\subsection{FedQTN model}
 In this section, we will focus on how medical AI can leverage the potential of quantum federated learning. 
To facilitate collaborative training with high-accuracy recognition and enhanced privacy in healthcare institutions, we focused on optimizing quantum tensor networks (QTNs) for medical imaging applications in the federated learning environment.

\begin{figure*}[!ht]
	\centering
	\includegraphics[scale=0.48]{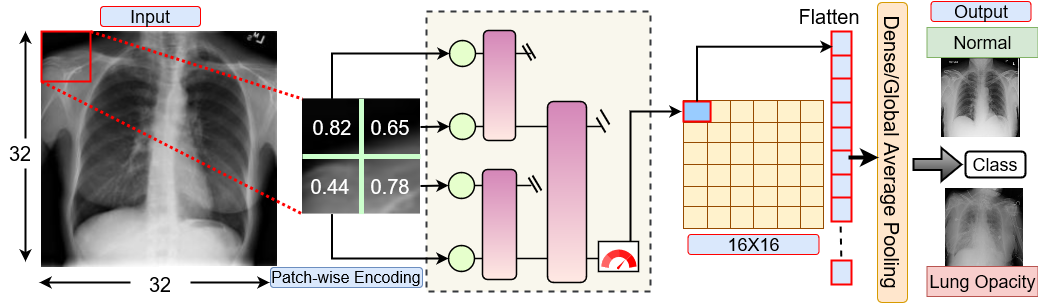}
	\caption{\textbf{Training the quantum tensor network model (TTN) locally on the client side}. The first step is to create patches of an input image, which are then encoded into a 4-qubit quantum circuit after flattening. Next, the outcomes of each patch from the quantum circuit are concatenated, followed by either dense layer or global average pooling operation to classify the image into either the normal or lung opacity class.  }
 \label{localmodel}
\end{figure*}

\subsection{Patch-based encoding}
 Before describing the federated learning approach based on quantum tensor architectures employed in this study,  it is important to elucidate the process of encoding classical data into the quantum circuit. 
 Consider a medical dataset $D=\{(x^i, y^i)\}_{i=1}^{n}$, included a N-dimensional input vectors $x^i$ and their corresponding labels $y^i \in \{0,1\}$. The first step is to encode the classical input vectors into the quantum state for classification on a quantum computer.
 It plays a crucial role in the performance of data-driven quantum machine learning algorithms. 
 Instead of encoding two-dimensional data like images in a straightforward fashion using a higher number of qubits, we used the concept of a patch-based qubit encoding procedure, as shown in Fig~\ref{encoding2}(a).
 We split the whole image $x_i$ into patches and encode each element of a patch in the amplitude of a single qubit. 
 The approach captures local entanglement more efficiently and relies exclusively on single-qubit rotations, but demands more storage capacity. 
 Initially, we normalize the pixels of each patch to lie in [0, $\pi/2$].
 Then, we flattened each \textit{n}-dimensional patch ($p_0, ..., p_{n-1}$) and mapped each pixel $p_j$ to the corresponding quantum superposition state independently.
\begin{equation}
    \ket{\psi_j}=\Big[cos\Big(\dfrac{\pi}{2} p_{j}\Big)\ket{0}+sin\Big(\dfrac{\pi}{2} p_{j}\Big)\ket{1}\Big]
\end{equation}

The final encoded image is a product state of these \textit{n}-dimensional pixel vectors quantum state $\ket{\psi}$. 
Now, the quantum dataset is represented as $D=\{(\psi^i, y^i)\}_{i=1}^{n}$.

\subsection{Local models}
 To perform model aggregation in the FedQTN framework, we considered a set of hospitals/clients $H$ participating in a global quantum tensor network model training. 
Each hospital $\{h_i\}_{i=1}^{H}$ uses its local quantum database $D_i$ containing patient medical images / samples $\{d_i\}_{i=1}^n$ to train its local model.
Local models are constructed on the basis of quantum circuit architectures such as MPS, TTN, and MERA. 
The low-bond dimension tensor networks can be mapped to quantum circuits using two-qubit unitary parameterization, as shown in Fig.~\ref{tensornetworks}.
However, multi-qubit gates are needed for higher bond dimensions between the tensor nodes.

The first quantum tensor network is inspired by a one-dimensional structure, i.e., MPS, in which the bond dimension two is visualized as a series of two-qubit unitary gates arranged in a staircase pattern.
Each tensor within the classical tensor network is obtained by contracting the initial two-qubit state with a subsequent two-qubit gate. 
The more complex hierarchical structures like TTN and MERA are considered. 
An example of local training of FedTTN on the client side is illustrated in Fig~\ref{localmodel}.
Each quantum circuit consists of a parameter-dependent unitary transformation $\mathcal{U}(\theta)$ acting on an input quantum state.
For the unitary transformation $\mathcal{U}(\theta)$, we have used a simple unitary gate and a strongly entangled unitary gate, as shown in Fig~\ref{encoding2}(b-c). 
In a simple unitary block, two single-qubit $R_y$ rotations, followed by the CNOT gate between the control and target qubits, and $R_y$ rotations.
In a strongly entangled unitary gate, each layer consists of two single-qubit $R_y$ rotations, followed by a forward CNOT and a reverse CNOT operation.
The application of the two-qubit unitary is iterated until only one qubit remains.
Subsequently, a quantum measurement operator ($\hat{O}$) is employed on the remaining qubit to predict the input state label as

\begin{equation}
    \tilde{y}_i=\braket{\psi_i|\hat{U}_{Q}^{\intercal}\mathcal{U}_i(\theta_i)^{\intercal}\hat{O}\mathcal{U}_i(\theta_i)\hat{U}_{Q}|\psi_i}
\end{equation}

\noindent  where $\hat{U}_{Q}(\{\mathcal{U}_i\})$ is a quantum circuit and $\theta_i$ is a set of parameters to define $\mathcal{U}_i(\theta_i)$. 
A specific qubit is measured ($\hat{O}$) through a basic Pauli-Z measurement operation. Following the procedure, each patch is encoded in a QTN and the results are concatenated.
Later, the global averaging pooling operation takes the average of the entire input feature map for final predictions. 
In this study, we used a binary cross-entropy (BCE) loss function to minimize the error between predictions $\tilde{y}_i$ and true class labels $y_i$, and an adaptive moment estimate (Adam) optimizer to update the classical parameters based on the cost function's gradient. 
The local loss function in the training database $D_i$ containing $\{d_i\}_{i=1}^n$ samples in the hospital ($h\in H$) can be represented as

\begin{equation}
    \mathcal{L}(\theta_h)=-\dfrac{1}{d_n} \sum^{d_i}_{i=1}{[y_i*log(\tilde{y}_i)+(1-y_i)*log(1-\tilde{y}_i)]}
\end{equation}

At each communication round (\textit{r}), the local model of each hospital generates updated parameters $\theta^{r}_h$ using a local ADAM optimizer with a learning rate $\eta_c$ as

\begin{equation}
    \theta^{r}_h=\theta^{r}_s-\eta_c \triangledown_{\theta_{r}^{h}}\mathcal{L}(\theta^{r}_{h}) 
\end{equation}

\noindent where $\mathcal{L}(\theta^{r}_{h})$ are the parameters of local model at hospital (\textit{h}) after applying the $\hat{U}_{Q}(\{\mathcal{U}_i\})$ quantum circuit, and $\theta^{r}_s$ denotes the global model parameters received at round (\textit{r}).
Overall, the objective is to collectively train the local QTN models so that each generalizes effectively on the different data distributions.

\subsection{Global model}
The global model is responsible for coordinating and aggregating the local models from each hospital. 
The aggregation process ensures that the knowledge gained by individual hospitals is combined to improve the overall QTN global model.
Upon completion of local training for communication round (\textit{r}), each hospital or client communicates its local model parameters ($\theta_h$) to the global server, maintaining the privacy of the local data within the hospitals that generated it.
Finally, the aggregation of all local model updates takes place using a federated averaging algorithm as
\begin{equation}
    \theta^r_{s} \leftarrow \theta^{r}_{h} - \eta_s \dfrac{1}{H}  \sum_{i=1}^{H} \theta^{r+1}_{h}
\end{equation}. The goal is to achieve an optimal global FedQTN model by minimizing the global loss function with a server's learning rate $\eta_s$. 
The global model broadcasts the updated parameters to all participants. 
This iterative process adjusts the aggregated model through the local training procedure across hospitals until the quantum model reaches a specified learning accuracy.

\subsection*{Experimental hardware and software}
\label{details}
We implemented the experiments on the Anvil system at Purdue University through allocation CIS240105 from the Advanced Cyberinfrastructure Coordination Ecosystem: Services \&
Support (ACCESS) program \cite{45}.
 Regarding software, the experiments were performed using TensorFlow 2.7.0 \cite{51} in Python 2.8.

\subsection*{Data availability}
 The RSNA chest X-ray that supports the findings of this study is available at https://www.rsna.org/rsnai/ai-image-challenge/rsna-pneumonia-detection-challenge-2018. The NIH chest X-ray dataset is publicly available at https://nihcc.app.box.com/v/ChestXray-NIHCC. The ADNI MRI-scan database can be accessed at https://adni.loni.usc.edu/  upon registration. The Kidney CT-scan data is publicly available at https://www.kaggle.com/datasets/nazmul0087/ct-kidney-dataset-normal-cyst-tumor-and-stone.

\section*{ACKNOWLEDGEMENTS}
DBN was supported by the NASA Academic Mission Services, Contract No. NNA16BD14C. 
DBN and AB acknowledge the support of the startup grant of the Davidson School of Chemical Engineering at Purdue University.

\bibliography{bibl}

\end{document}